\documentclass[twocolumn,secnumarabic,amssymb, nobibnotes, aps, prb]{revtex4}
\usepackage{graphicx}
\usepackage{amsmath}
\usepackage{epsfig,rotating,graphicx,graphics,color,bm}

\def \be{\begin{equation}}
\def \ee{\end{equation}}

\newcommand{\pade}[2]{\ensuremath{\frac{\partial #1}{\partial #2}}}

\begin{document}

\title{A simple vision of current induced spin torque in domain walls}
\author{A. Vanhaverbeke}%
\author{M. Viret}
\affiliation{Service de physique de l'\'etat condens\'e (CNRS URA 2464),\\
CEA Saclay F-91191 Gif-sur-Yvette cedex, France\\}
\date{september 20, 2006}%

\begin{abstract}
The effective spin pressure induced by an electric current on a domain wall in a ferromagnet is determined using a simple classical model, which allows us to extend previous theories to arbitrary domain wall widths. In particular, the role of spatially non-uniform components of the torques are analysed in detail. We find that the effect of the current is mainly to distort the domain wall which should enhance de-pinning. We also find that in the limit of thin domain walls the current-induced pressure changes sign.
\end{abstract}

\maketitle


Traditionally, spin electronics has been dealing with resistance changes induced by different magnetic configurations, like those encountered in 'Giant Magneto-Resistance' and 'Tunneling Magneto-Resistance' devices. This has led to important developments in information technology applied in computer read-heads and the more recent magnetic memories (or MRAMs) for high density information storage. For the latter, one has to be able to read and write the magnetic information. Writing is presently achieved by local application of a magnetic field using current lines crossing on each individual cell. This procedure is the main source for cross-talk between magnetic elements because of field leakages which can influence neighbouring memories. Thus, it would be most useful to switch the magnetic configurations directly with a current flowing in the spin-valve. The relevant effect is called spin-torque and it has been predicted by Slonczewski\cite{sclonczewski-jmmm-159} and Berger \cite{berger-prb-54} and observed in the late 90'~\cite{albert-apl-77}. It is now well-known that magnetization can be reversibly switched in ferromagnetic-normal metal-ferromagnetic trilayers by a large current crossing the interfaces. Besides, it appears that this effect can also produce microwave-frequency oscillations of the thin layer~\cite{kiselev-nature-425}, possibly allowing coherent microwave generation by an assembly of nanopillars~\cite{slavin-prb-72}. An alternative way to control magnetic configurations is to move domain walls (DWs) with a current as first proposed by Berger~\cite{berger-jap-55} and demonstrated a long time ago \cite{freitas-jap-57}. Recently, theoretical studies~\cite{zhang-prl-93} and numerical calculations~\cite{thiaville-epl-69} have shown that a large part of the current-induced torque does not push the DWs and the pressure only originates from spin-flip events relaxing the conduction electrons' spins.

We propose here a simple picture of the relevant physics involved in the generation of the different components of the torque in domain walls. The basis for it lies in understanding the evolution of the spin of a conduction electron as it crosses a domain wall and its reaction on the local moment. The relevant interaction is that between localized (the local magnetisation) and delocalized (conduction) electrons which can be expressed with the \emph{s-d} Hamiltonian:
$H_{s-d}=-J_{ex}\vec{s}\cdot\vec{S}$ where $J_{ex}$ is the exchange interaction, $<\vec{S}>/S=-\vec{M}/M_s$ refers to localized spins, and $\vec{s}$ to the conduction electrons spins. The exchange interaction splits the conduction electrons in two populations with spins parallel (up) or antiparallel (down) to the local moments. A current generates a plane wave of electrons whose wave functions can be expressed as a spinnor with two components, up and down, travelling with different wave vectors. The question is then to analyse how these two components evolve when the electrons are forced to cross a region where localized moments change direction in space, i.e. a domain wall. The proper way to do this is to write Schr\"odinger's equation and match wave functions and their derivatives at the borders of the DW. Solutions of the problem are complex for any particular value of the domain wall width, but analytical solutions can be easily found for the two limits of an abrupt, or a very long domain wall (see for example ref. \cite{waintal-epl-65}).
There is another way of treating the problem which is to consider conduction electrons as free particles entering a region in space where a local field changes direction (the DW). Their spin evolution is then obtained by writing the Landau-Lifshitz equation. This is simpler, but one has to overcome conceptual problems linked with the nature of the electrons crossing, which are band particles. These two visions of the problem are actually equivalent in the limit where the amount of reflected wave can be neglected \cite{waintal-epl-65}. Hence, this is valid for a DW width much greater than the Fermi wavelength \cite{cabrera-pssb-61}. Several works in that field (addressing more particularly DW resistance) have chosen either the 'wave' approach \cite{tatara-prl-92,brataas-prb-60,li-prl-92} or the 'particle' one \cite{berger-jpcs-35,viret-prb-53}. Because in the wave approach one has to use perturbation theory, spin torques have been calculated in the limits of wide, or very narrow domain walls. Here, we propose to use the simplicity of the 'particle' approximation to describe the spin evolution in the domain walls and to extend previous results to intermediate values of the DW width.

\bigskip
\textbf{Classical model:}

If both magnetization and electron magnetic moment ($\vec{\mu}=-g\mu_B\vec{s}$) are considered classical vectors, the spin dynamics obeys the basic precession equation:

\begin{equation}
\frac{d\vec{\mu}}{dt}=\;\frac{J_{ex}S}{\hbar}\; \vec{m}\times\vec{\mu}
\label{eq:spinprecess}
\end{equation} 
where $\vec{m}=\vec{M}/M_s$ is the unitary vector of magnetization. This is similar to the classical Landau-Lifshitz equation with the external field replaced by the exchange field, which reflects the difference between the s-d Hamiltonian considered here and the simple Zeeman one.

\begin{figure}[htbp]
\includegraphics[width=0.3\textwidth]{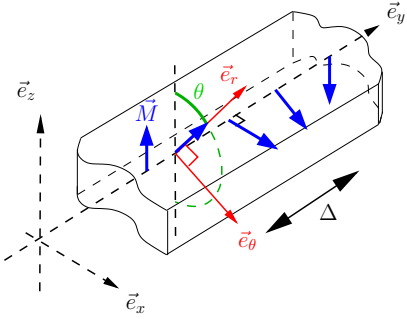}
\caption{Bloch domain wall : conventions}
\label{fig:bloch}
\end{figure}

We consider here the frame moving with the electron crossing the DW, in which the magnetization varies continuously in time. For simplicity, the magnetisation rotates in the wall plane (Bloch type) but the following derivation would equally apply to a rotation around any axis (as long as the system can be considered one-dimensional). The moving frame is defined by the three vectors ($\vec{e_r}$,$\vec{e_\theta}$,$\vec{e_y}$) as shown in figure~\ref{fig:bloch}. The evolution of the electron magnetic moment is thus simply described by:

\begin{equation}
\frac{d\vec{\mu}}{dt}=\left(
\begin{array}{c}
\dot{\mu_r}-\dot{\theta}\mu_{\theta} \\
\dot{\mu_{\theta}}+\dot{\theta}\mu_r \\
\dot{\mu_y}
\end{array}\right)
=\frac{SJ_{ex}}{\hbar}\left(\begin{array}{c}1\\0\\0\end{array}\right)\times\left(\begin{array}{c}\mu_r\\\mu_\theta\\\mu_y\end{array}\right)
\end{equation}

Defining $\tau_{ex}=\hbar/SJ_{ex}$ we get:

\begin{equation}
\left\{
\begin{array}{l}
\dot{\mu_r}-\dot{\theta}\mu_{\theta}=0\\
\dot{\mu_\theta}+\dot{\theta}\mu_r=-\frac{\mu_y}{\tau_{ex}}\\
\dot{\mu_y}=\frac{\mu_{\theta}}{\tau_{ex}}
\end{array}
\right.
\label{eq:mmevolution}
\end{equation}

\begin{figure}[htbp]
	\centering
\includegraphics[width=0.23\textwidth]{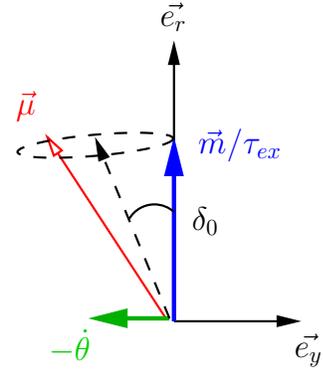}
	\caption{Evolution of the conduction electron magnetic moment in the frame rotating with the magnetization in the domain wall. The precession of $\vec{\mu}$ is not around the local magnetization but around the effective field direction $\vec{m}/\tau_{ex}-\dot{\theta}\vec{e}_y$}
	\label{fig:rotref}
\end{figure}

As the conduction electrons cross the DW, their spins precess with the so called Larmor period. It is convenient to consider first a long DW (where $\dot{\theta}\tau_{ex}\gg1$) with a linear variation of the magnetization angle $\theta$ (the second derivative of the angle is neglected $\ddot{\theta}=0$, and the electrons spin remains mainly aligned with the magnetization), which leads to the simplified equations: 

\begin{eqnarray}
\ddot{\mu_{\theta}}+\frac{1}{\tau_{ex}^2}\mu_{\theta}  & = & 0 \nonumber\\
\ddot{\mu_y}+\frac{1}{\tau_{ex}^2}\mu_y & = &-\frac{\dot{\theta}}{\tau_{ex}}\frac{g \mu_B}{2}
\end{eqnarray}

In fact the magnetic moment $\vec{\mu}$ is precessing around an effective field (fig.~\ref{fig:rotref}) whose direction is given by the sum of the magnetisation and its rotation vector: $\vec{m}/\tau_{ex}-\dot{\theta}\vec{e}_y$. Every precession, the electron's magnetic moment lags from the magnetization and then catches up \cite{viret-prb-53}. As a consequence, the average over time of the mistracking between $\vec{\mu}$ and the magnetization is not zero as it lies on the effective field:

\begin{equation}
<\vec{\mu}>=\frac{g\mu_B}{2}\left(\begin{array}{c}
1\\
0\\
-\dot{\theta}\tau_{ex}
\end{array}\right)
\label{newequation}
\end{equation}

This is a crucial point as it explains why, even if we consider a large number of Larmor oscillations of $\vec{\mu}$ in the wall, the resulting effect on the magnetization will not be averaged to zero. Because at any time during DW crossing the total magnetic moment has to be preserved, a torque is induced on the magnetization by each electron (per unit volume):

\begin{equation}
\frac{\delta\vec{M}}{\delta t}=\frac{1}{\tau_{ex}}\vec{\mu}\times\left(\begin{array}{c}1\\0\\0\end{array}\right)\\
\end{equation}

This part is adiabatic as there is no energy loss in the process. 
The resulting torque can be decomposed into a constant and a periodic term \cite{waintal-epl-65}.  
For long DWs, the periodic part averages to zero and the constant \emph{spin torque} term:

\begin{equation}
\frac{\delta\vec{M}}{\delta t}\Big|_{st} = 
\frac{1}{\tau_{ex}}\left(
\begin{array}{c}\frac{g \mu_B}{2}\\0\\<\mu_y>\end{array}
\right)\times\left(
\begin{array}{c}1\\0\\0\end{array}
\right)
 =  -\frac{g\mu_B}{2}\dot{\theta}\vec{e_{\theta}}
\end{equation}

is not in a proper direction to move the DW~\cite{thiaville-epl-69}. Recently, it was shown that the term responsible for current induced DW motion is due to a non-adiabatic part of the torque~\cite{zhang-prl-93}. In order to account for it in our classical model, it is necessary to include the spin-flip relaxation time:
\begin{equation}
\frac{d\vec{\mu}}{dt}=-\frac{1}{\tau_{ex}} \vec{m}\times\vec{\mu}-\frac{1}{\tau_{sf}}( \vec{\mu}-\vec{\mu}_{eq})
\label{spinprecessrelax}
\end{equation}
where $\vec{\mu}_{eq}=g\mu_B/2 \vec{e}_r$ is aligned with the magnetization. 
In usual ferromagnetic materials, the spin-flip relaxation time $\tau_{sf}$ is large compared to the Larmor period $T_L=2\pi\tau_{ex}$, and the effect can be considered a perturbation. Then the average mistracking of the electron magnetic moment $\mu_y$ as derived in eq.~\ref{newequation} can be kept. The induced effect is given by the reaction torque of this spin-flip related term on the local magnetization:

\begin{equation}
\frac{\delta\vec{M}}{\delta t}\Big|_{sf} = \frac{1}{\tau_{sf}}<\mu_y>\vec{e}_y = -\frac{g\mu_B}{2}\dot{\theta}\frac{\tau_{ex}}{\tau_{sf}}\vec{e_{y}}
\end{equation}

Under an electric current density $\vec{j}=j\vec{e}_y$, and choosing to introduce the magnetization gradient in the wall we then get for the two torques:

\begin{eqnarray}
\frac{d\vec{M}}{dt}\Big|_{st}								&=& 
\frac{jP}{e}\frac{g\mu_B}{2}\pade{\vec{m}}{y} \\ 														
\frac{d\vec{M}}{dt}\Big|_{sf}								&=&	\frac{jP}{e}\frac{g\mu_B}{2}\frac{\tau_{ex}}{\tau_{sf}}\left(\vec{m}\times\pade{\vec{m}}{y}\right) 
\end{eqnarray}

Where the polarization $P$ is added as a prefactor to account for the partial polarization of the charge carriers. So the adiabatic spin torque accounted by various authors~\cite{berger-jap-71,berger-prb-33,thiaville-apl-95,waintal-epl-65} can be explained as the reaction from the precession of the conduction electrons magnetic moments around the effective field. The second term originating from the spin flip scattering has an amplitude reduced by a factor $\tau_{ex}/\tau_{sf}$ typically around 1/30.
When writing the micromagnetic equations of motion of the DW, one can see that only this (small) second term applies a pressure that pushes the wall~\cite{thiaville-apl-95,zhang-prl-93}. This can also be seen in the following manner: from the Landau-Lifshitz equation it can be seen that the precession of a conduction electrons' spin around the magnetization is equivalent to a field oriented along the rotation vector acting on the magnetization. Because the main spin evolution of electrons crossing the wall is a rotation following the DW magnetization, the equivalent field they generate is directed perpendicular to the plane of the DW and hence to the magnetization of the domains. As a result, this does not push the wall in any specific direction but only induces a canting of the wall magnetization. On the other hand, spin-flip processes tend to align the conduction electrons spins with the local magnetization. The effective field this generates is perpendicular to both the local magnetization and the initial direction of the electron spin as the event occurs. Considering average quantities during DW crossing, the local magnetization lies globally along its direction in the middle of the wall and the electron spin lags behind along the direction of rotation of M (that given by $\dot{\theta}$). Thus, on average, this effective field is perpendicular to both vectors, i.e. it is along the direction of the domains magnetization. Therefore, this is applying a pressure which tends to push the DW.

Hence, our classical model leads to deformation and pressure expressions which are consistent with the results of semi-classical theories. The advantage of our simple formalism is twofold. 
It is now possible to study in detail the spatial evolution of the torque along the width of the DW, and it also makes possible to explore the torque for wall widths of the order of the precession length: $\Delta \simeq \lambda_L$. Indeed, the only essential approximation of the model is that the reflected part of the plane electron wave impinging on the DW can be neglected. This imposes DW widths much larger than the Fermi wavelength of the electrons~\cite{cabrera-pssb-61,waintal-epl-65}, which is not a very stringent condition since the latter is around 3\AA~in metals.


\bigskip
\textbf{Numerical calculations for thin walls:}

\begin{figure}[htbp]
\centering
\includegraphics[width=0.42\textwidth]{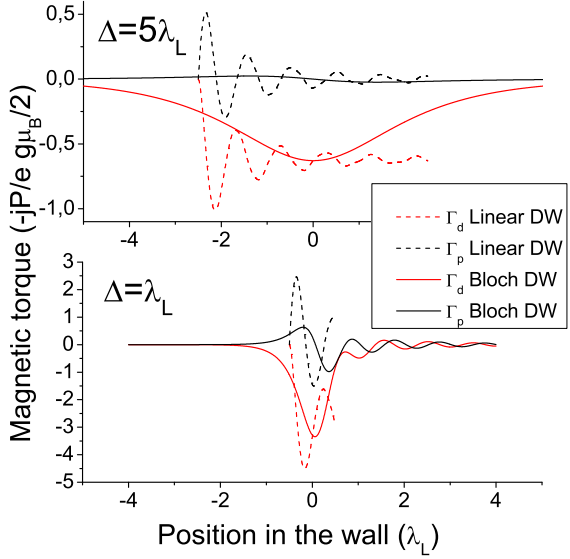}
\caption{"Distortion" and "Pressure" torque parameters $\Gamma_d$ and $\Gamma_p$ in the domain wall under a current density $j$ with $\Delta=5\lambda_L$ (top curve) and $\Delta=\lambda_L$ (bottom) and $\frac{\lambda_{ex}}{\lambda_{sf}}=1/30$.}
\label{fig:torqueinwall}
\end{figure}

\begin{figure}[htbp]
	\centering
		\includegraphics[width=0.4\textwidth]{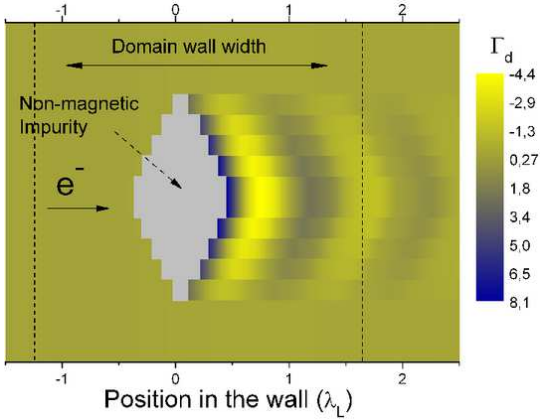}
	\caption{Numerical simulation of the effect of a non-magnetic impurity on the torque $v_p$, for a Bloch DW of width $\Delta=2.5\lambda_L$. The figure is a superposition of 1D simulations for the varying width of the non-magnetic material, considering it does not affect the domain wall profile.}
	\label{fig:defect}
\end{figure}

We have numerically calculated the current-induced torques exerted locally on the wall from eq.~\ref{spinprecessrelax} without the simplifications associated to large domain walls. Pressure and deformation terms are labelled respectively $\Gamma_p$ and $\Gamma_d$:
\begin{eqnarray}
\Gamma_d(y)	&=& \frac{jP}{e}\left<\frac{\mu_{y}(y,v)}{\tau_{ex}}+\frac{\mu_{\theta}(y,v)}{\tau_{sf}}\right>_{v}\\
\Gamma_p(y)	&=& -\frac{jP}{e}	\left<\frac{\mu_{\theta}(y,v)}{\tau_{ex}}+\frac{\mu_y(y,v)}{\tau_{sf}}\right>_{v}
\end{eqnarray}  
These are also averaged on the different directions the Fermi velocity can take on the Fermi sphere. This induces decoherence of the torques as electrons travelling with different components of their velocity on the direction perpendicular to the DW, $v_y$, have different Larmor precession lengths. The resulting effect is a damping of the averaged spin precession (hence the oscillating part of both torques) after a few Larmor periods $2\pi\tau_{ex}$. This allows us to extend the work of Waintal and Viret~\cite{waintal-epl-65} who derived the periodic torque but did not include the spin-flip scattering, which lead to a zero average of $\Gamma_p$.
Figure~\ref{fig:torqueinwall} presents numerical computations of the torques along different walls: linear and Bloch type, with $\Delta=5\lambda_L$ or  $\Delta=\lambda_L$ (with $\lambda_L=\frac{v_F\tau_{ex}}{2\pi}$)  and $\lambda_{sf}=50\lambda_{ex}$. An obvious result is that the exact DW shape is important for the torques as also pointed out by Xiao et al. \cite{xiao-condmat-0601172}. For a linear wall (dashed line) where $\dot{\theta}$ is discontinuous, oscillations are enhanced compared to those in a Bloch wall (solid line). Thus, in the pure Bloch walls of bulk materials, the periodic torque is small but in short linear walls, like these expected in constrictions \cite{Kazantseva-prl-94}, periodic components are significant. In real systems, where DWs are often pinned on defects or impurities, large torque oscillations can also be expected because of the abrupt perturbations defects have on the local magnetization. This is schematically shown in fig. \ref{fig:defect} for a Bloch wall pinned on a non-magnetic impurity. Beyond the simplicity of the chosen system, the result of these large torque oscillations should be an efficient depinning of the walls.

Let us now turn to the average torque values on the domain wall width $\Delta$. To allow for a straightforward comparison with ref.~\cite{li-prl-92} we introduce the two velocity-like parameters $v_d$ and $v_p$ corresponding to the torques exerted on the magnetization, respectively in the $\vec{e}_{\theta}$ and $\vec{e}_y$ directions: 
\begin{equation}
v_{d,p}	= \frac{1}{|\pade{m}{y}|}	\frac{1}{\Delta} \int_{y} \Gamma_{d,p}(y) dy 
\end{equation}

\begin{figure}[htbp]
	\centering
		\includegraphics[width=0.35\textwidth]{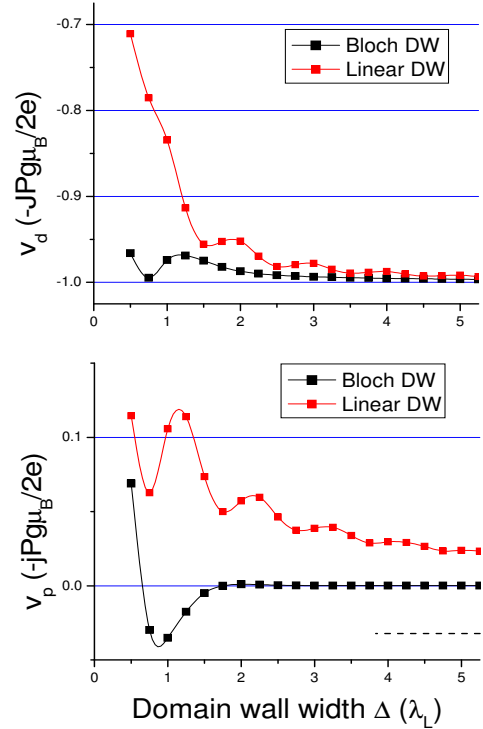}
	\caption{Averaged distortion ($v_d$) and pressure ($v_p$) as a function of the domain wall width $\Delta$ for linear and Bloch walls. The component applying pressure($v_p$) changes sign and increases steeply for thin walls (the asymptotic value for wide linear walls is represented by the dashed line).}
	\label{fig:torquesav}
\end{figure}

The dependence of the relevant quantities $v_d$ and $v_p$ with $\Delta$ is presented in fig.~\ref{fig:torquesav}. Asymptotic values for large domain walls ($\Delta\gg\lambda_L$) are different for linear and Bloch walls. This reflects the different relevant lengthscales associated to each wall: in a linear wall, because of the abrupt border of the DW, the Larmor precession is maximum and it becomes the only relevant lengthscale whereas for a smooth Bloch wall, the periodic part of the torque is smeared and the wall width becomes more relevant. Hence, when the DW width is larger than the spin diffusion length, the torque decreases because the relaxation reduces the average mistracking, and hence the pressure torque. As a result, the asyptotic values for $v_p$ in fig.~\ref{fig:torquesav} are $-\frac{\lambda_{ex}}{\lambda_{sf}}$ for linear walls and zero for Bloch walls.

For thin domain walls with width close to the Larmor length $\lambda_L$, the oscillatory torque neither tends to zero nor is it washed out by the average on the Fermi sphere. Interestingly, as $\Delta$ is reduced, the $v_p$ term changes sign and becomes much larger than its value for wide walls: $\frac{jPg\mu_B}{2e}\frac{\tau_{ex}}{\tau_{sf}}$. This is at first sight surprising, but can be understood as follows: when the electrons first enter the DW, the magnetization rotation generates a tilt in the effective field around which the conduction electrons spins precess. Hence, their first period takes them away from the local magnetization and the corresponding torque is opposite to that generated by spin-flip scattering. It is also much larger because at first all the electrons are precessing in that same way. In very thin DWs, this effect dominates. Thus, in appropriate systems, the effective torque could be an order of magnitude larger than that in conventional domain walls. This is particularly relevant for ferromagnetic materials with very large anisotropy, where the predicted domain wall width is only a few nanometers, which is close to the Larmor precession length $\lambda_L$. One could also imagine making walls thin in nanometer-sized constrictions, or increasing $\lambda_L$ as in magnetic semiconductors. This might actually explain results in GaMnAs, where DWs have been shown to move under much lower current densities \cite{yamanouchi-nature-428} corresponding to an efficiency of 30\% for the spin torque (an order of magnitude higher than in NiFe). This would be the case in our calculation for $\Delta\simeq2\lambda_L$, which is reasonable in GaMnAs because of a relatively weak exchange and large anisotropy. Interestingly, for thin walls, conduction electrons are injected in the magnetic domain beyond the DW with their spin significantly misaligned. Spin flip events will eventually realign the spins thus leaving some angular momentum within a spin diffusion length from the DW. It is likely that this will result in large magnon emission in the domains. This effect is beyond the scope of this paper, but one can conjecture that for a large current, the magnetization beyond the DW could be destabilized.

In summary, the model we develop here allows us to study the evolution of the conduction electrons spins as they cross a domain wall. For long walls, the average torque is mainly due to a constant term coming from the global spin rotation while following the magnetization direction. This tends to distort the DW without pushing it. The only contribution applying a pressure is that due to spin-flip scattering events which amounts to a few percent of the total torque. For thinner walls, the contribution from the periodic torque, which depends on the exact shape of the DW, becomes important. The total torque can be significantly enhanced and it is even found to change sign for very thin walls. Thus, we predict that for domain walls of width close to the Larmor precession length, the current induced spin pressure is an order of magnitude higher and in a direction opposite to that for large DWs.

We acknowledge financial support from the 'Ministère de la Recherche et de l'Education' through the 'ACI Nanosciences NR216 PARCOUR'. 

\bibliographystyle{apsrev}

\end{document}